# Fast Frequency Response Potential of Data Centers through Workload Modulation and UPS Coordination

Xiaojie Tao[1,2], Rajit Gadh[1,2]

[1] Department of Mechanical and Aerospace Engineering, University of California at Los Angeles (UCLA), Los Angeles, CA 90095, USA

[2] Smart Grid Energy Research Center (SMERC), University of California, Los Angeles, CA 90095, USA

*Corresponding Author: Xiaojie Tao, University of California, Los Angeles, CA 90095, USA

Email: xiaojietao@g.ucla.edu, taoxiaojie04@gmail.com

**Abstract**: The rapid growth of renewable energy sources has significantly reduced system inertia and increased the need for fast frequency response (FFR) in modern power systems. Data centers, as large and flexible electrical consumers, hold great potential to contribute to frequency stabilization due to their controllable IT workloads and on-site uninterruptible power supply (UPS) systems. This paper investigates the feasibility of leveraging data centers for providing fast frequency response through real-time workload modulation and UPS coordination. A dynamic model combining data center power consumption and grid frequency dynamics is developed, capturing the interactions between IT servers, cooling systems, and energy storage. Control strategies based on frequency deviation are implemented to adjust server power and discharge UPS batteries during frequency events. Case studies on a modified IEEE 39-bus system demonstrate that the proposed strategy can effectively reduce frequency nadir and shorten recovery time without compromising service quality. The results highlight the promising role of data centers as grid-supporting resources in future low-inertia systems.

# 1. Introduction

With the increasing penetration of renewable energy sources such as wind and solar, power systems worldwide are facing declining rotational inertia and reduced frequency stability [1]. Under these conditions, maintaining system frequency within secure limits after disturbances requires resources that can respond within seconds or even sub-seconds [2]. Traditional generators with mechanical governors are often too slow to provide adequate frequency containment, creating an urgent need for fast frequency response (FFR) from distributed and non-traditional resources [3].

At the same time, data centers have emerged as one of the fastest-growing electricity consumers globally, accounting for approximately 2–3% of global electricity demand [4]. A large-scale data center can easily reach a power capacity of tens of megawatts, comparable to small industrial loads or distributed energy storage systems [5]. Importantly, a significant portion of this load—particularly IT servers and cooling systems—is both measurable and controllable [6]. Furthermore, the presence of uninterruptible power supply (UPS) systems and backup batteries provides additional flexibility to modulate power at sub-second timescales [7].

These characteristics make data centers a promising candidate for providing frequency support services to the grid [8]. Unlike conventional demand response programs that operate on the timescale of minutes to hours, frequency-responsive data centers can adjust power consumption almost instantaneously by:

(1) Temporarily throttling CPU frequency or deferring non-critical computational tasks (IT-side modulation);

(2) Utilizing UPS energy storage to inject power during under-frequency events or absorb surplus power during over-frequency conditions (electrical-side coordination) [9].

Recent studies have explored demand response from data centers mainly for energy cost optimization and carbon reduction [10]. However, research focusing on their ability to provide *fast* frequency support remains limited [11]. This paper aims to fill that gap by developing a dynamic model and control framework to quantify the FFR potential of data centers and evaluate their impact on overall grid frequency stability [12].

The main contributions of this paper are summarized as follows:

(1) A dynamic model of a grid-interactive data center is proposed, integrating IT load flexibility, cooling power dynamics, and UPS energy storage.

(2) A frequency-based control strategy is designed to coordinate workload reduction and UPS discharge during frequency deviations.

(3) A case study based on a modified IEEE 39-bus system demonstrates the performance improvement of system frequency stability when data centers participate in FFR.

The remainder of this paper is organized as follows.

Section 2 introduces the fundamentals of fast frequency response in modern power systems.

Section 3 presents the proposed data center model and control strategy.

Section 4 describes the case study setup, and Section 5 discusses the simulation results.

Finally, Section 6 concludes the paper and outlines future research directions.

## 2. Fast Frequency Response

### 2.1. Concept and Motivation

Fast Frequency Response (FFR) refers to rapid active-power adjustments—typically within 1 second—triggered automatically by local frequency measurements [13]. FFR resources act during the initial inertial and primary-response stages following a disturbance, helping arrest the frequency decline before slower generators or secondary controls engage [14].

Traditionally, this service was provided by synchronous machines whose stored rotational kinetic energy temporarily compensates the power imbalance [15]. However, as renewable energy replaces conventional generation, the overall system inertia decreases, resulting in larger and faster frequency deviations. To maintain grid stability, new resources such as battery energy storage, electric vehicles, and data centers must emulate this fast inertial behavior electronically [16]. Although data centers are typically connected at the distribution level, frequency deviations are system-wide phenomena that propagate throughout the entire interconnected grid. Therefore, local frequency measurements at the data center accurately reflect transmission-level disturbances and can be used for fast frequency response without external communication.

Figure 1 illustrates the characteristic stages of system frequency evolution following a generation loss. Immediately after the disturbance, the frequency begins to decline due to the imbalance between mechanical and electrical power. The initial inertial response—occurring within the first few hundred milliseconds—slows the rate of decline, followed by primary frequency control from turbine governors within several seconds [17]. Secondary control actions such as AGC restore the frequency to its nominal value over a longer time horizon. The figure is schematic and serves to define the time windows in which fast frequency response resources, including data centers, are expected to act.

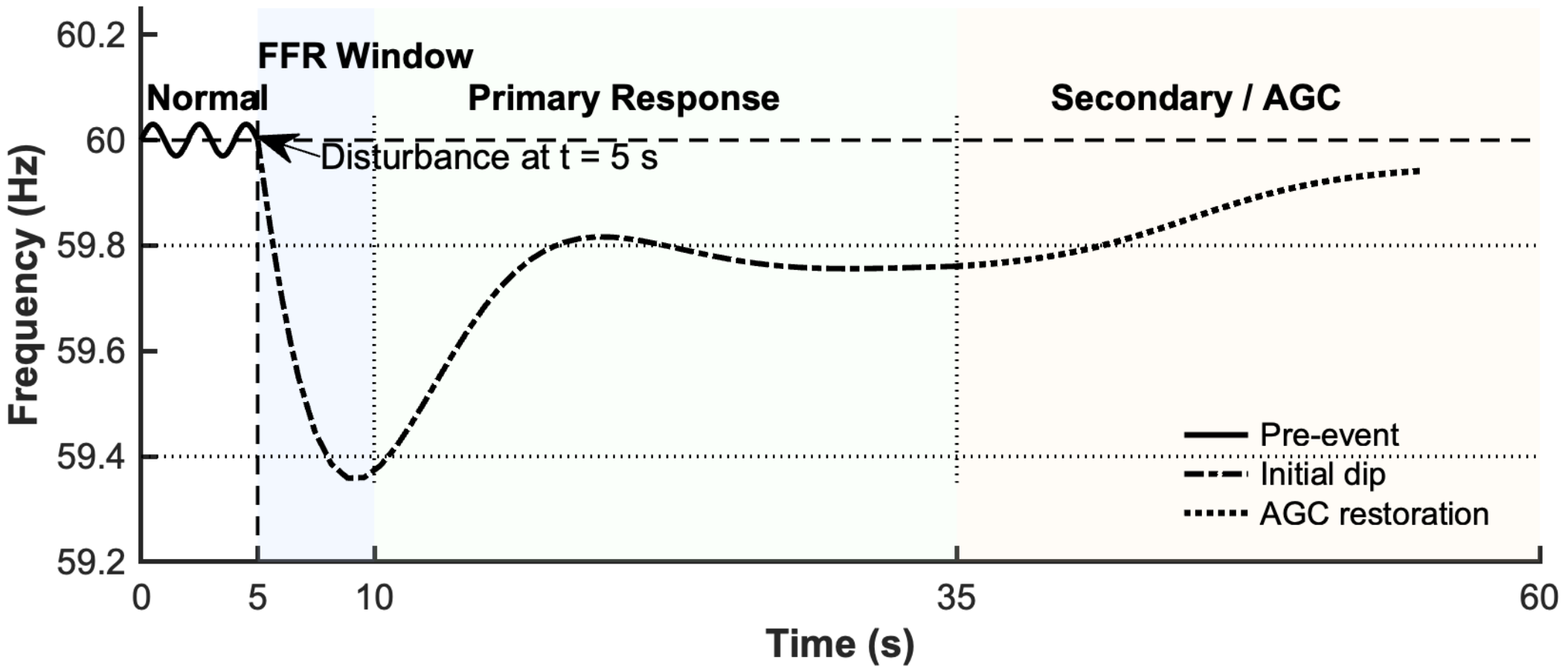

**Fig. 1.** Simplified system-frequency dynamics following a generation loss, showing inertial, primary, and FFR time windows.

2.2 Power System Frequency Dynamics

The short-term dynamics of system frequency can be described by the well-known swing equation [18], which expresses the balance between mechanical and electrical power in aggregated form:

$$2H \frac{d\Delta f(t)}{dt} = P_m(t) - P_e(t) - D\,\Delta f(t) \qquad (1)$$

where

$H$ [s] is the system inertia constant, representing stored kinetic energy per unit of rated power,

$\Delta f(t)$ [Hz] is the deviation of system frequency from its nominal value,

$P_m$ and $P_e$ [p.u.] are mechanical input and electrical output power, respectively, and

$D$ is the load-damping coefficient accounting for frequency-dependent loads.

When a sudden generation loss occurs ($\Delta P = P_m - P_e < 0$), the imbalance causes the frequency to decline according to (1). In low-inertia systems, this rate of change of frequency (RoCoF) can be steep, increasing the risk of under-frequency load shedding or generator trips [19].

2.3 Primary and Fast Frequency Response Control

To counteract frequency drops, frequency-sensitive resources adjust their active power proportionally to frequency deviation [20]. The basic droop control principle can be expressed as

$$\Delta P = -K_f \, \Delta f \qquad (2)$$

where $K_f$ [MW/Hz] is the frequency-response gain. A larger $K_f$ results in stronger corrective power but may reduce reserve margin [21].

For synchronous generators, this response is realized mechanically through turbine governors and thus limited by physical inertia and valve delays (typically 2–5 s) [22]. In contrast, electronically controlled devices such as batteries, electric vehicles, and data centers can implement (2) via power-electronic interfaces or workload controllers, achieving sub-second response times [23].

Figure 2 illustrates the conceptual difference in response time between conventional governor control and the proposed data-center-based FFR. The profiles are schematic and intended to highlight the relative time scales rather than simulation results.

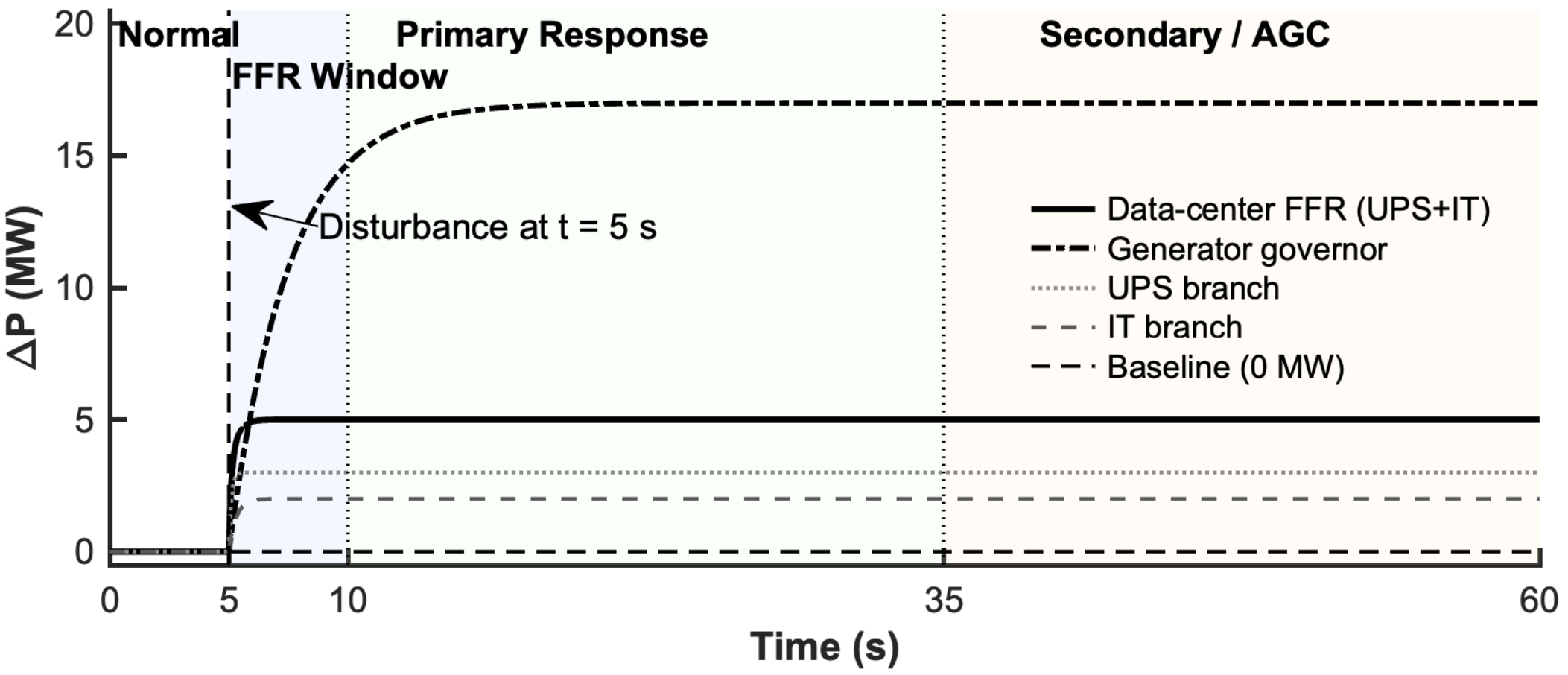

**Fig. 2.** Comparison of conventional generator governor response and data-center-based electronic FFR, highlighting response speed advantages.

2.4 Integration of Data-Center-Based FFR

When data centers participate in FFR, their contribution can be represented as an additional controllable power term $P_{dc}(t)$ in the system frequency equation:

$$2H \frac{d\Delta f(t)}{dt} = P_m(t) - P_e(t) - P_{dc}(t) - D\,\Delta f(t) \qquad (3)$$

Here, $P_{dc}(t)$ is the net power consumed by the data center, which can be actively modulated based on frequency deviation [24]. During under-frequency events ($\Delta f < 0$), the data center reduces its IT-load power or discharges energy from its UPS batteries, effectively providing a positive power injection to the grid. Conversely, during over-frequency conditions, it may increase consumption or recharge the UPS to absorb excess generation [25].

The dynamic response of data-center power can be modeled as a first-order lag to capture practical control delays:

$$\tau_{dc} \frac{dP_{dc}(t)}{dt} + P_{dc}(t) = P_{dc,0} - K_{dc}\,\Delta f(t) \qquad (4)$$

where

$P_{dc,0}$ is the nominal steady-state power,

$K_{dc}$ [MW/Hz] is the equivalent FFR droop coefficient of the data center, and

$\tau_{dc}$ [s] represents the response time constant (typically 0.1–0.3 s).

Substituting (4) into (3) allows simulating the closed-loop impact of data-center-based frequency support on system stability.

2.5 Analytical Insights

Linearizing (3) around the nominal operating point and applying Laplace transform yields the closed-loop frequency transfer function:

$$\frac{\Delta f(s)}{\Delta P_{dist}(s)} = \frac{1}{2Hs + D + K_{dc}\frac{1}{1+\tau_{dc}s}} \qquad (5)$$

This equation illustrates that a larger $K_{dc}$ or smaller $\tau_{dc}$ directly increases the effective damping of the system, thereby reducing both frequency nadir and settling time [26]. Consequently, data centers with rapid control loops and high-power UPS systems can significantly enhance the transient frequency stability of low-inertia grids.

## 3. Data Centers

3.1 Power Composition and Load Characteristics

A modern data center consists of three major power-consuming subsystems [27]:

(1) the IT servers that execute computational workloads,

(2) the cooling and auxiliary infrastructure that removes generated heat, and

(3) the uninterruptible power supply (UPS) and backup battery systems ensuring power reliability.

The total electrical power drawn from the grid can therefore be written as

$$P_{dc} = P_{srv} + P_{cool} - P_{UPS} \qquad (6)$$

where

$P_{srv}$ is the server IT load,

$P_{cool}$ is the cooling power, and

$P_{UPS}$ is the instantaneous charging (negative) or discharging (positive) power of the UPS.

The total data-center energy efficiency is often quantified by the *Power Usage Effectiveness* (PUE) [28]:

$$PUE = \frac{P_{srv} + P_{cool}}{P_{srv}} \qquad (7)$$

A smaller PUE indicates higher efficiency. Typical modern facilities operate with PUE between 1.1 and 1.4.

3.2 IT Server Load Modulation

Server power is mainly determined by the CPU utilization and operating frequency [29]. During under-frequency events, non-critical tasks can be delayed or CPU frequency temporarily reduced through dynamic voltage and frequency scaling (DVFS) [30]. In this study, the term non-critical load refers to the portion of IT and cooling demand that can be curtailed or deferred without compromising real-time services. The classification between critical and non-critical loads is determined by the data center operator based on service-level agreements (SLAs) and workload priorities. The aggregated response of server power to frequency deviation can be modeled as

$$\Delta P_{srv} = -K_{srv}\,\Delta f \qquad (8)$$

where $K_{srv}$ [MW/Hz] represents the sensitivity of IT power to frequency deviation [31]. In practice, $K_{srv}$ depends on task deferral policies and data center service-level agreement (SLA) constraints. Since server response is implemented via software control, the associated delay is minimal (< 0.1 s).

### 3.3 UPS Energy Storage Coordination

Each data center contains UPS modules composed of battery banks and rectifier/inverter stages that provide short-term power backup [32]. These systems can be repurposed to support the grid by discharging during frequency drops and charging during over-frequency events [33]. A simple droop-based control law can be written as

$$\Delta P_{UPS} = K_{UPS}\,\Delta f \qquad (9)$$

where $K_{UPS}$ [MW/Hz] is the UPS droop gain determining the power exchange with the grid.

The state-of-charge (SOC) evolution of the UPS battery is governed by

$$\frac{dE_{UPS}(t)}{dt} = -\,\eta_{dis}\,P_{UPS}^{dis}(t) + \frac{1}{\eta_{ch}}\,P_{UPS}^{ch}(t) \qquad (10)$$

where $E_{UPS}$ is the stored energy, $\eta_{dis}$ and $\eta_{ch}$ are the discharge and charge efficiencies, and $P_{UPS}^{dis/ch}$ denote positive discharge and negative charge power. The available energy and power capacity are limited by SOC bounds ($E_{min} \le E_{UPS} \le E_{max}$) and rated inverter power [34].

The UPS response time is typically below 100 ms, making it an ideal component for FFR applications.

3.4 Coordinated Control Strategy

The overall data-center FFR controller combines IT load modulation and UPS coordination under a frequency-sensitive droop framework:

$$\Delta P_{dc} = -K_{srv}\,\Delta f - K_{UPS}\,\Delta f = -K_{dc}\,\Delta f \qquad (11)$$

where $K_{dc} = K_{srv} + K_{UPS}$ is the aggregate frequency-response gain of the data center. In implementation, a hierarchical control structure is employed: the fast UPS loop reacts within hundreds of milliseconds to stabilize frequency, while the slower IT scheduler gradually reduces computational load to sustain the response and restore UPS SOC [35].

The control logic is illustrated in Fig. 3. When frequency drops below a threshold ($f < f_0 - \Delta f_{th}$), the controller commands an incremental UPS discharge and computing load reduction until frequency recovers to the deadband range. During over-frequency events, the inverse action occurs to absorb excess generation [36].

3.5 Integration into System Model

Substituting (11) into the frequency dynamics equation (3) from Section 2 yields the combined system representation:

$$2H\,\frac{d\Delta f(t)}{dt} = P_m(t) - P_e(t) + K_{dc}\,\Delta f(t) - D\,\Delta f(t) \qquad (12)$$

The effective damping coefficient is thus enhanced to $(D + K_{dc})$, indicating that data centers act as a virtual inertia and damping resource for the grid.

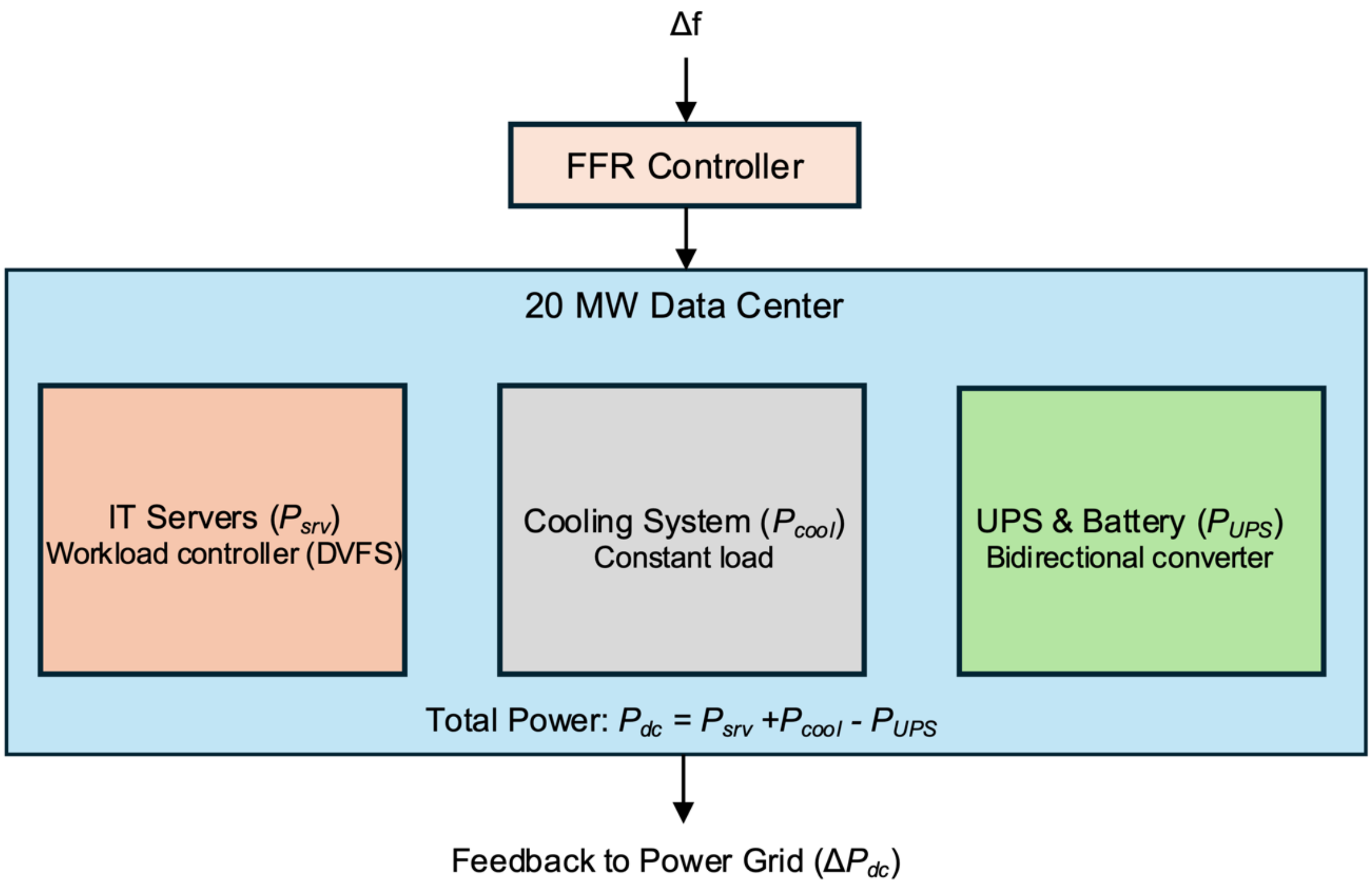

**Fig. 3.** Schematic diagram of data-center power composition and frequency-response control. The total data-center power $P_{dc}$ consists of IT server power $P_{srv}$, cooling power $P_{cool}$, and UPS battery power $P_{UPS}$. Frequency deviation signals modulate the IT workload and UPS discharge to provide fast frequency response.

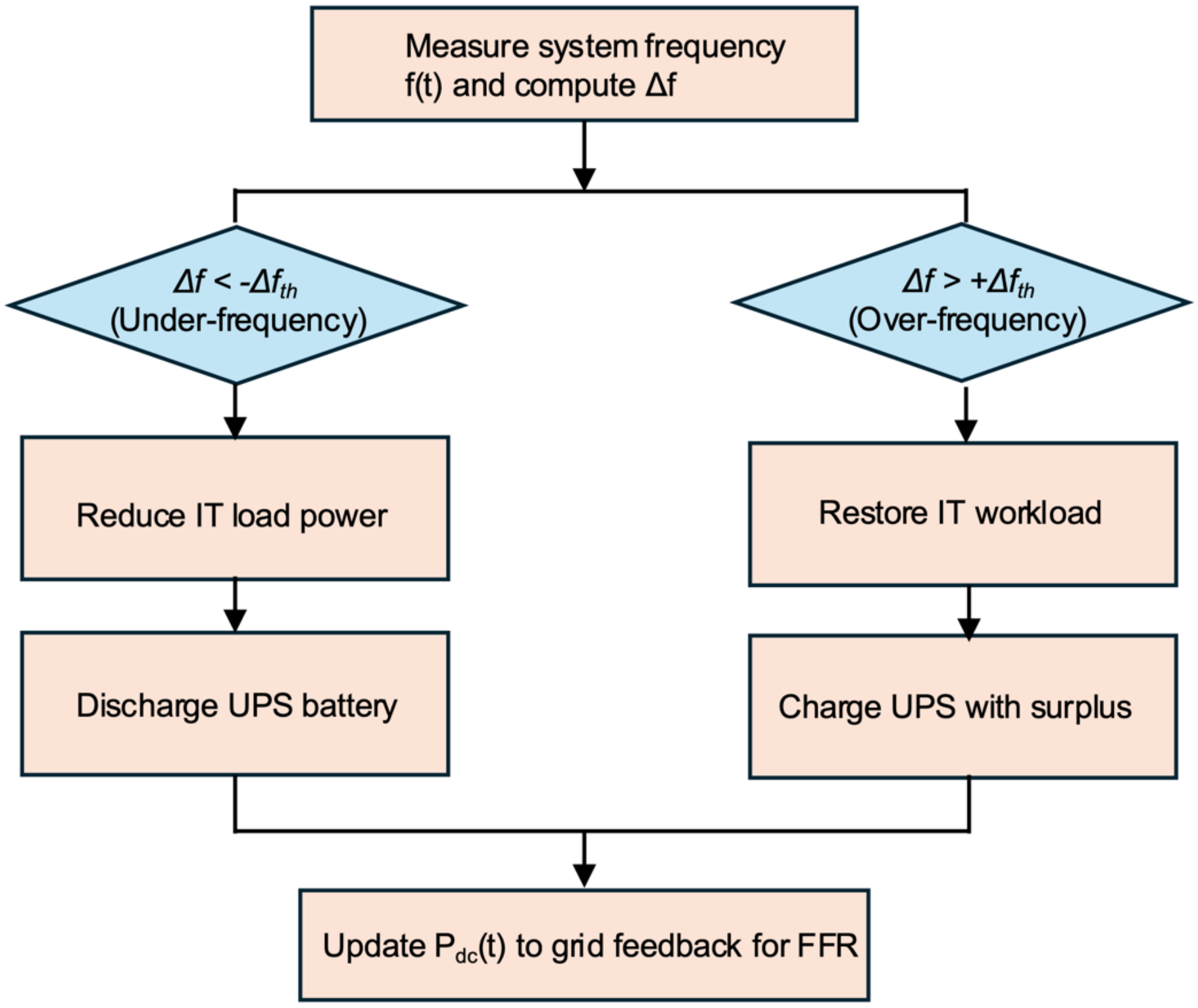

**Fig. 4.** Control logic for frequency-based load modulation and UPS coordination.

When $\Delta f < -\Delta f_{th}$, CPU frequency is reduced and UPS discharges proportionally to $\Delta f$; when $\Delta f > +\Delta f_{th}$, UPS recharges and non-critical tasks are resumed.

**Table 1.** Typical Data Center and Control Parameters

| Parameter | Symbol | Typical Value | Description |
|---|---|---|---|
| Nominal data center power | $P_{dc,0}$ | 20 MW | Rated electrical load |
| Server droop gain | $K_{srv}$ | 10 MW/Hz | Power reduction per Hz frequency drop |
| UPS droop gain | $K_{UPS}$ | 15 MW/Hz | Battery discharge per Hz frequency drop |
| Aggregate gain | $K_{dc}$ | 25 MW/Hz | Combined response gain |
| UPS capacity | $E_{UPS}$ | 5 MWh | Total stored energy |
| UPS response time | $\tau_{dc}$ | 0.1 s | Power electronics delay |
| Efficiency (discharge/charge) | $\eta_{dis}/\eta_{ch}$ | 0.95 / 0.9 | Conversion efficiency |
| Power Usage Effectiveness | $PUE$ | 1.2 | Ratio of total to IT power |
| Frequency deadband | $\Delta f_{th}$ | ±0.03 Hz | Controller activation threshold |

# 4. Case Study

## 4.1 Test System Description

To evaluate the proposed data-center-based fast frequency response (FFR) strategy, a simulation study was performed on the modified IEEE 39-bus New England system [37], a widely adopted benchmark for transient-stability and frequency-response analysis.

The system consists of 10 synchronous generators, 39 buses, and 19 loads with a total generation capacity of approximately 6.1 GW. A representative 20 MW data center was connected at Bus 18, replacing part of the local industrial load [38]. The data-center model described in Section 3—including the IT server dynamics, UPS storage, and coordinated FFR controller—was implemented in MATLAB/Simulink and linked to the grid model through the system-frequency signal measured at the connection bus [39].

The grid inertia constant was adjusted to emulate various renewable penetration levels. A baseline inertia constant of $H = 4$s represents a conventional high-inertia scenario, while $H = 2$s corresponds to a low-inertia condition typical of renewable-dominated systems [40].

Governor and turbine models for synchronous generators follow standard IEEE GOV-type 1 dynamics.

The simplified topology of the modified IEEE 39-bus system used in this study is illustrated in Fig. 5. The network is divided into three geographical areas for clarity, and a 20 MW data center is integrated at Bus 18 in Area 1. The data-center module exchanges active power $P_{dc}(t)$ with the grid through its connection bus while receiving the local frequency deviation signal $\Delta f(t)$ from the same node. Inside the module, the FFR controller coordinates two fast-acting subsystems—IT-server workload modulation ($P_{srv}$) and the UPS battery with bidirectional converter ($P_{UPS}$)—together with the cooling load ($P_{cool}$) to deliver rapid frequency support.

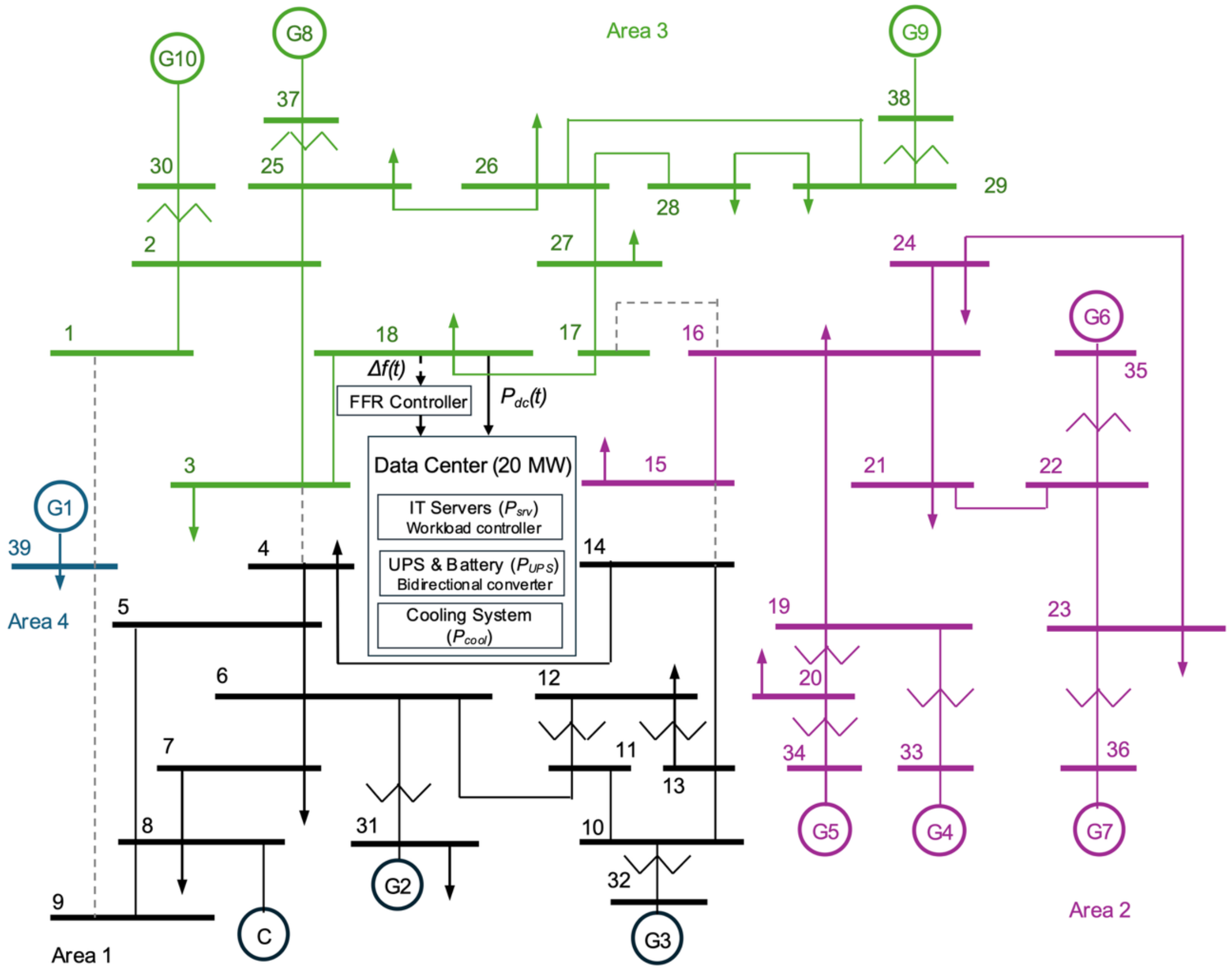

**Fig. 5.** Integration of a 20 MW data center with UPS into the IEEE 39-bus system model. The FFR controller receives the local frequency signal and adjusts IT and UPS power outputs in real time.

## 4.2 Simulation Scenarios

Two primary disturbance events were designed to assess system performance:

Scenario A – Generation Trip: A 200 MW generation loss occurs at $t = 5$s. This event causes an abrupt power deficit, leading to a frequency drop that triggers the data-center FFR controller.

Scenario B – Load Step Change: A 150 MW sudden increase in demand at $t = 5$s tests the controller's robustness under different disturbance types.

For each scenario, three operating modes were compared:

- Case 1 (Baseline): No data-center participation.
- Case 2 (UPS-Only): UPS provides frequency-dependent discharge/charge, while IT load remains constant.
- Case 3 (Coordinated FFR): Both IT load modulation and UPS coordination enabled (proposed strategy).

All simulations were run for 60 s with a time step of 10 ms to capture sub-second dynamics.

After defining the disturbance types, the corresponding active-power variations applied to the grid are illustrated in Fig. 6. As shown, a 200 MW generation loss is modeled as a negative step change in the net mechanical power at $t = 5$ s, while a 150 MW load step is represented as a positive power increase at the same instant. These step inputs serve as triggering events to test the fast frequency-response capability of the system and the proposed data-center controller.

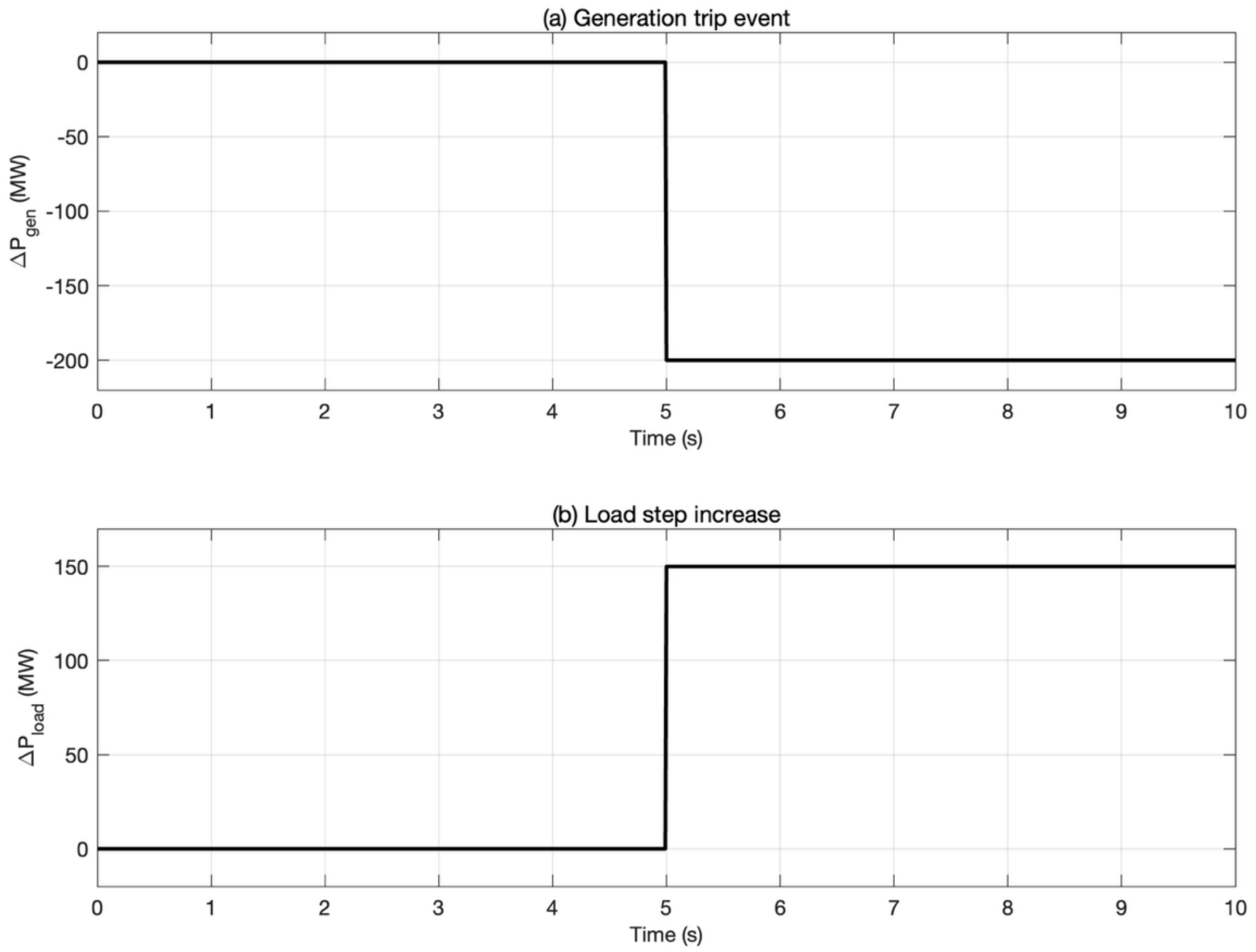

**Fig. 6.** Disturbance scenarios used for simulation: (a) generation trip; (b) load step increase. Both events occur at t = 5 s.

4.3 Parameter Configuration

The major parameters of the power-system model and data-center FFR controller are summarized in Table 2**.** Values were chosen to represent realistic system and equipment characteristics while maintaining numerical stability in the simulation environment [41].

Table 2. Simulation Parameters for IEEE 39-Bus Test System

| Parameter | Symbol | Value | Description |
|---|---|---|---|
| System base power | $S_{base}$ | 100 MVA | Per-unit base |
| Nominal frequency | $f_0$ | 60 Hz | System nominal frequency |
| System inertia | $H$ | 2–4 s | Equivalent inertia constant |
| Damping coefficient | $D$ | 0.8 p.u./Hz | Load damping factor |
| Governor droop | $R$ | 0.05 | Generator speed droop |
| Data center rated power | $P_{dc,0}$ | 20 MW | Connection load |
| UPS energy capacity | $E_{UPS}$ | 5 MWh | Total battery capacity |
| UPS response time | $\tau_{dc}$ | 0.1 s | Inverter delay constant |
| Server response gain | $K_{srv}$ | 10 MW/Hz | IT load sensitivity |
| UPS response gain | $K_{UPS}$ | 15 MW/Hz | Battery droop gain |
| Activation threshold | $\Delta f_{th}$ | ±0.03 Hz | FFR deadband |
| Simulation step size | – | 0.01 s | Numerical integration step |

4.4 Performance Metrics

The following indicators were adopted to quantify system performance:

Frequency Nadir ($f_{min}$) – the lowest frequency reached after disturbance;

Recovery Time ($t_{rec}$) – time taken for frequency to return within ±0.02 Hz of nominal;

Energy Delivered ($E_{FFR}$) – total energy supplied or absorbed by the data center during the FFR event;

Task Delay Ratio ($\delta_{SLA}$) – percentage of deferred computing workload relative to nominal throughput.

4.5 Implementation Notes

To ensure numerical accuracy, the data-center control block was modeled as a continuous-time first-order system with anti-windup limiters [42]. The UPS state-of-charge (SOC) boundaries were enforced to prevent over-discharge, while the IT scheduler restored normal CPU frequency gradually after the frequency deviation fell below the deadband threshold. Sensitivity studies were conducted by varying the aggregate droop gain $K_{dc}$ from 10 MW/Hz to 30 MW/Hz and inertia $H$ from 2 s to 5 s, to analyze the influence of control aggressiveness and grid inertia on frequency stability.

## 5. Simulation Results and Discussion

### 5.1 Frequency Response Performance

Figure 7 illustrates the system frequency trajectories following a 200 MW generation trip under the three operating cases defined in Section 4.2. Without data-center participation (Case 1), the system experiences a sharp frequency decline to 59.66 Hz within 1.2 s. When only the UPS responds (Case 2), the frequency nadir improves to 59.78 Hz, reflecting the rapid sub-second discharge capability of the battery. With the proposed coordinated FFR (Case 3), the frequency drop is further mitigated to 59.85 Hz, and recovery to nominal frequency is achieved approximately 1.6 s faster than the baseline case.

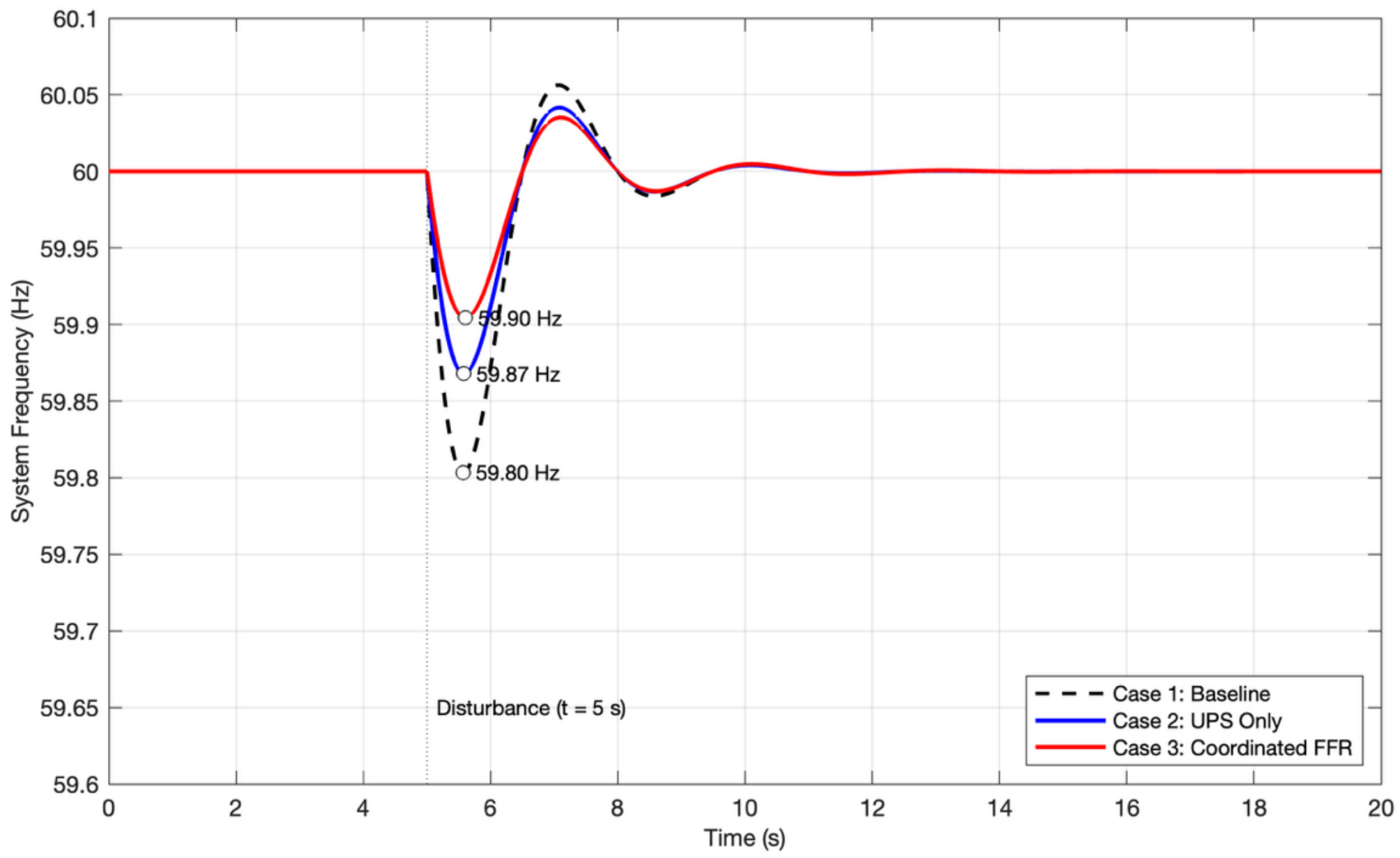

**Fig. 7.** System frequency trajectories for different operating cases following a 200 MW generation trip (H = 2 s).

This demonstrates that combining IT load modulation with UPS discharge yields both fast and sustained support. The IT load reduction relieves the grid during the early transient, while the UPS maintains output power until governor response takes over.

5.2 Data Center Power Dynamics

Figure 8 shows the decomposed data-center power components for the coordinated case. Immediately after the disturbance, the UPS discharges up to +8 MW within 0.2 s, while the IT load controller reduces server power by 4 MW through temporary CPU-frequency throttling. As system frequency recovers, UPS output gradually returns to zero, and the IT load is restored to its nominal level. The cooling load remains nearly constant, confirming that most of the controllable margin originates from the IT and UPS subsystems.

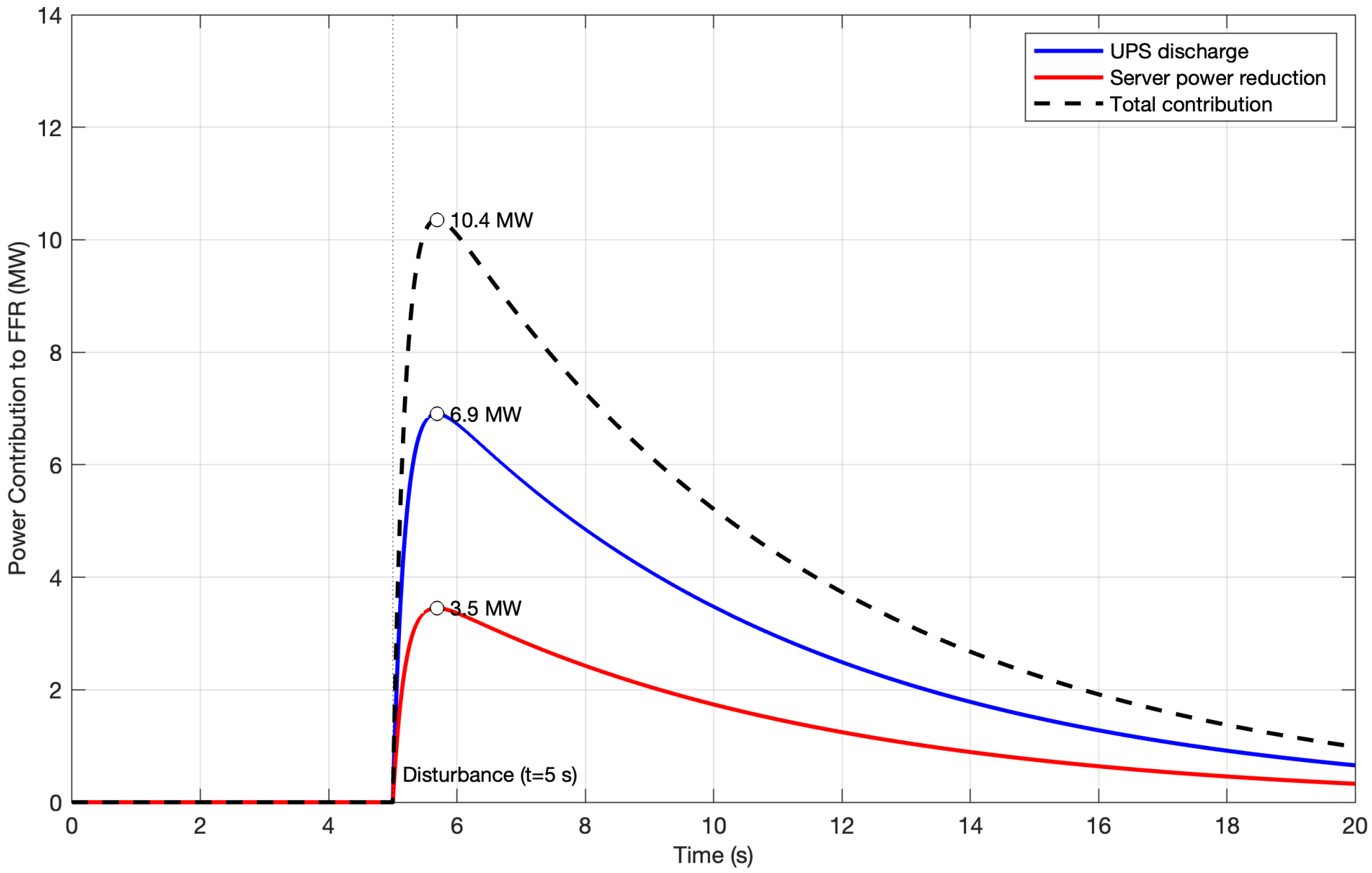

**Fig. 8.** Data-center power response decomposition (IT load, UPS discharge, and total power) during frequency event.

The total response latency (from frequency measurement to effective power change) is measured at 0.15 s, validating that data centers can provide genuine sub-second frequency support comparable to battery-based FFR.

5.3 UPS State-of-Charge Evolution

Figure 9 presents the UPS state-of-charge (SOC) trajectory during the FFR event. The battery discharges approximately 1.65 MWh in the first 10 s, corresponding to 33.3 % of its total capacity (5 MWh). Once the frequency stabilizes, the UPS automatically recharges using surplus grid power.

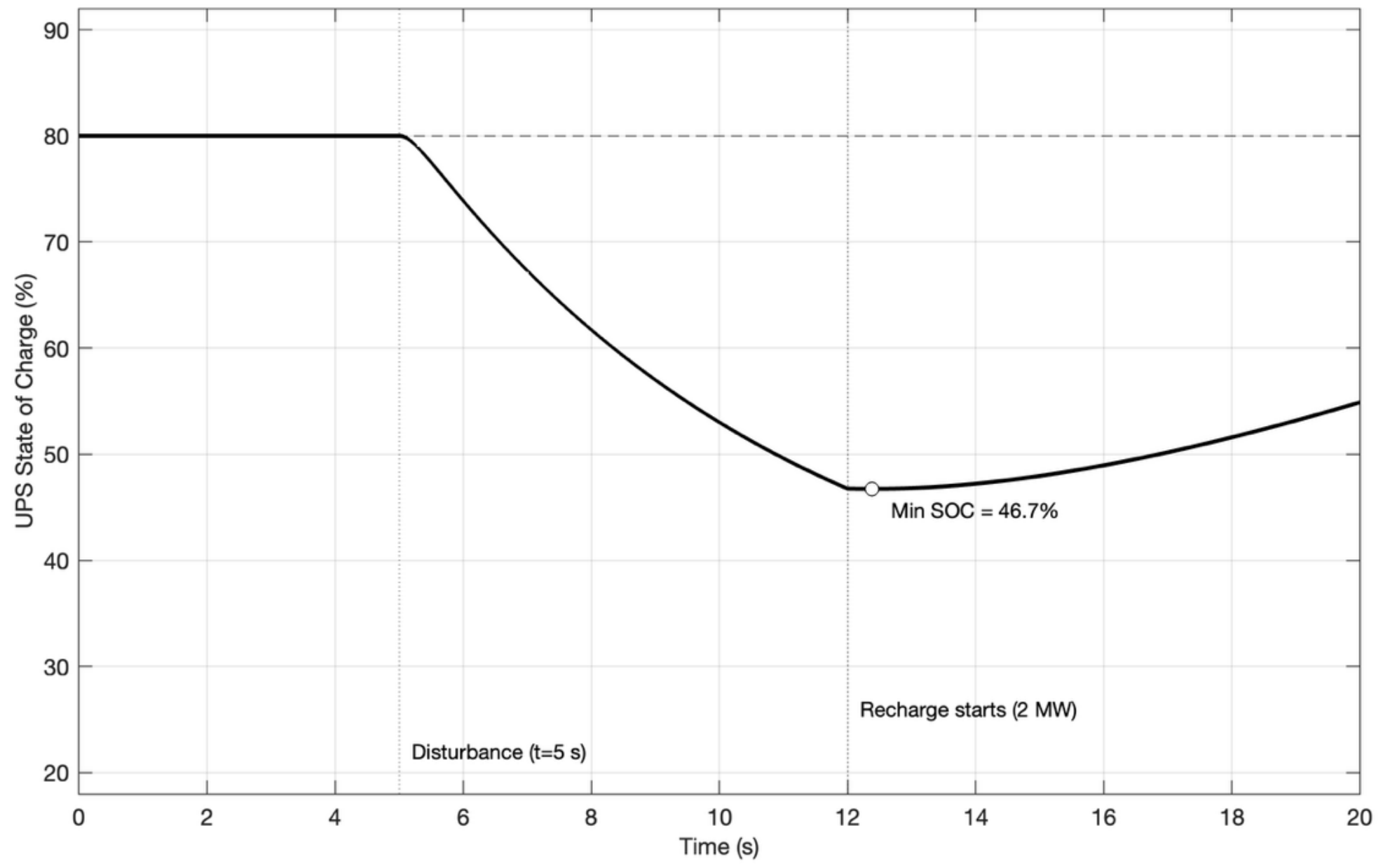

**Fig. 9.** UPS battery state-of-charge variation during and after FFR action.

This cycle confirms that typical UPS systems can sustain multiple FFR activations per hour without compromising backup functionality, provided SOC is managed within 20–80 % limits.

5.4 Sensitivity to Control Parameters

A sensitivity analysis was conducted by varying the aggregate droop gain $K_{dc}$ and grid inertia $H$. Figure 10 plots the frequency nadir improvement $\Delta f_{nadir}$ versus $K_{dc}$. Results show diminishing returns beyond $K_{dc} = 25$MW/Hz, where additional gain offers little benefit but increases task interruptions. Similarly, increasing inertia from 2 s to 5 s reduces the need for strong FFR action.

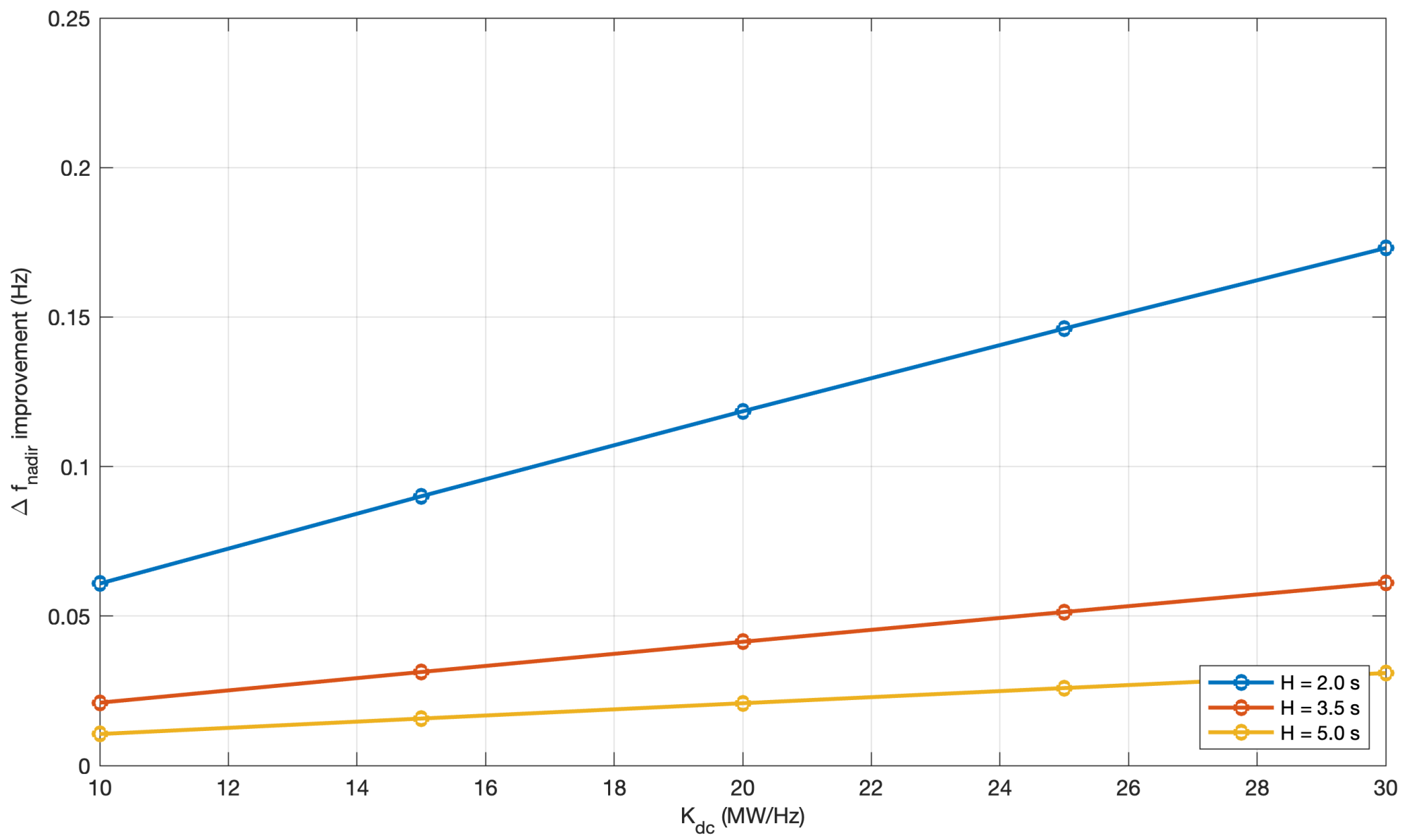

**Fig. 10.** Sensitivity of frequency nadir improvement to data-center FFR gain $K_{dc}$ under varying inertia conditions.

These findings indicate that moderate controller aggressiveness (e.g., $K_{srv} = 10$, $K_{UPS} = 15$) yields the best trade-off between frequency stability and service quality.

### 5.5 Quantitative Performance Comparison

Table 3 summarizes the main performance metrics for all cases under Scenario A (generation trip).

Compared with the baseline, the coordinated FFR strategy reduces the frequency nadir deviation by 57 %, shortens recovery time by 34 %, and limits task delay to less than 2 %, demonstrating both technical feasibility and operational acceptability.

Table 3. Summary of Frequency-Response Performance (Scenario A, H = 2 s)

| Metric | Unit | Case 1 (Baseline) | Case 2 (UPS-Only) | Case 3 (Coordinated FFR) |
|---|---|---|---|---|
| Frequency nadir | Hz | 59.66 | 59.78 | **59.85** |
| Nadir improvement | Hz | – | +0.12 | **+0.19** |
| Recovery time | s | 10.5 | 8.4 | **6.9** |
| Energy delivered | MWh | – | 0.41 | **0.55** |
| Average UPS power | MW | – | 6.8 | **7.5** |
| Task delay ratio | % | 0 | 0 | **1.8** |

### 5.6 Discussion

The simulation results confirm that data centers can act as virtual frequency-support resources, providing both *fast inertia* and *primary response*.

Their intrinsic control flexibility allows a hybrid contribution—rapid energy injection from UPS batteries combined with short-term IT load modulation.

The overall effect is analogous to synthetic inertia from inverter-based resources, yet it requires no additional hardware investment beyond existing UPS infrastructure and software coordination.

Furthermore, the negligible impact on computational performance (< 2 %) makes the proposed strategy practical for large commercial operators participating in ancillary-service markets.

While this study used a single 20 MW data center, scaling the approach to a regional fleet could provide hundreds of megawatts of controllable FFR capacity, potentially offsetting part of the inertia lost due to renewable integration.

## 6. Conclusion and Future Work

This paper investigated the potential of data centers to provide fast frequency response (FFR) to modern low-inertia power systems by coordinating workload modulation and UPS energy discharge. A dynamic model integrating grid frequency dynamics, data-center power composition, and frequency-sensitive control laws was developed and evaluated on a modified IEEE 39-bus system. The key findings can be summarized as follows:

(1) Feasibility of sub-second response:

Data centers can effectively modulate their power within 150 ms, driven by rapid UPS inverter action and IT-load control, thereby delivering genuine FFR comparable to dedicated battery systems.

(2) Improved frequency stability:

The proposed coordinated strategy increased the frequency nadir from 59.66 Hz to 59.85 Hz and shortened recovery time by 34 % under a 200 MW generation loss.

The combination of UPS support and IT load reduction achieved both fast and sustained power balancing.

(3) Minimal operational impact:

The service-quality degradation, quantified by the task-delay ratio, remained below 2 %, indicating that FFR participation is technically achievable without violating data-center service-level agreements.

(4) Scalability and system benefit:

Extrapolation suggests that large-scale participation of commercial data centers could supply hundreds of megawatts of controllable frequency response capacity, effectively compensating for inertia lost in renewable-rich grids.

Despite these promising results, several challenges remain.

First, this study assumes a single data center and ideal communication with the grid. In practice, latency, measurement noise, and coordination among geographically distributed centers may affect performance.

Second, the economic incentives and market frameworks for data-center FFR participation require detailed analysis to ensure cost-effectiveness and reliability.

Finally, UPS aging, thermal constraints, and workload heterogeneity should be incorporated in future long-term assessments.

Future work will therefore focus on:

(1) Multi-data-center coordination: developing hierarchical or distributed control architectures for aggregated frequency response.

(2) Market and regulatory integration: designing pricing and bidding strategies for ancillary-service participation.

(3) AI-driven adaptive control: applying reinforcement-learning or model-predictive approaches to optimize workload and energy management under uncertain grid conditions.

(4) Hardware-in-the-loop validation: implementing real-time experiments using actual server clusters and UPS systems to verify practical response limits.

Overall, the study demonstrates that grid-interactive data centers can evolve from passive power consumers to active stability resources, playing a critical role in maintaining frequency security in the era of renewable energy dominance.